
\message
{JNL.TEX version 0.92 as of 4/24/89.  Using CM fonts.}

\catcode`@=11
\expandafter\ifx\csname inp@t\endcsname\relax\let\inp@t=\input
\def\input#1 {\expandafter\ifx\csname #1IsLoaded\endcsname\relax
\inp@t#1%
\expandafter\def\csname #1IsLoaded\endcsname{(#1 was previously loaded)}
\else\message{\csname #1IsLoaded\endcsname}\fi}\fi
\catcode`@=12

\font\twelverm=cmr12			\font\twelvei=cmmi12
\font\twelvesy=cmsy10 scaled 1200	\font\twelveex=cmex10 scaled 1200
\font\twelvebf=cmbx12			\font\twelvesl=cmsl12
\font\twelvett=cmtt12			\font\twelveit=cmti12
\font\twelvesc=cmcsc10 scaled 1200	\font\twelvesf=cmss12
\skewchar\twelvei='177			\skewchar\twelvesy='60


\def\twelvepoint{\normalbaselineskip=12.4pt plus 0.1pt minus 0.1pt
  \abovedisplayskip 12.4pt plus 3pt minus 9pt
  \belowdisplayskip 12.4pt plus 3pt minus 9pt
  \abovedisplayshortskip 0pt plus 3pt
  \belowdisplayshortskip 7.2pt plus 3pt minus 4pt
  \smallskipamount=3.6pt plus1.2pt minus1.2pt
  \medskipamount=7.2pt plus2.4pt minus2.4pt
  \bigskipamount=14.4pt plus4.8pt minus4.8pt
  \def\rm{\fam0\twelverm}          \def\it{\fam\itfam\twelveit}%
  \def\sl{\fam\slfam\twelvesl}     \def\bf{\fam\bffam\twelvebf}%
  \def\mit{\fam 1}                 \def\cal{\fam 2}%
  \def\sc{\twelvesc}		   \def\tt{\twelvett}
  \def\sf{\twelvesf}
  \textfont0=\twelverm   \scriptfont0=\tenrm   \scriptscriptfont0=\sevenrm
  \textfont1=\twelvei    \scriptfont1=\teni    \scriptscriptfont1=\seveni
  \textfont2=\twelvesy   \scriptfont2=\tensy   \scriptscriptfont2=\sevensy
  \textfont3=\twelveex   \scriptfont3=\twelveex  \scriptscriptfont3=\twelveex
  \textfont\itfam=\twelveit
  \textfont\slfam=\twelvesl
  \textfont\bffam=\twelvebf \scriptfont\bffam=\tenbf
  \scriptscriptfont\bffam=\sevenbf
  \normalbaselines\rm}


\def\beginlinemode{\endmode
  \begingroup\parskip=0pt \obeylines\def\\{\par}\def\endmode{\par\endgroup}}
\def\beginparmode{\endmode
  \begingroup \def\endmode{\par\endgroup}}
\let\endmode=\par
{\obeylines\gdef\
{}}
\def\singlespace{\baselineskip=\normalbaselineskip}

\def\oneandahalfspace{\baselineskip=\normalbaselineskip
  \multiply\baselineskip by 3 \divide\baselineskip by 2}
\def\doublespace{\baselineskip=\normalbaselineskip \multiply\baselineskip by 2}

\newcount\firstpageno
\firstpageno=2
\footline={\ifnum\pageno<\firstpageno{\hfil}\else{\hfil\twelverm\folio\hfil}\fi}
\def\toppageno{\global\footline={\hfil}\global\headline
  ={\ifnum\pageno<\firstpageno{\hfil}\else{\hfil\twelverm\folio\hfil}\fi}}
\let\rawfootnote=\footnote		
\def\footnote#1#2{{\rm\singlespace\parindent=0pt\parskip=0pt
  \rawfootnote{#1}{#2\hfill\vrule height 0pt depth 6pt width 0pt}}}
\def\raggedcenter{\leftskip=4em plus 12em \rightskip=\leftskip
  \parindent=0pt \parfillskip=0pt \spaceskip=.3333em \xspaceskip=.5em
  \pretolerance=9999 \tolerance=9999
  \hyphenpenalty=9999 \exhyphenpenalty=9999 }


\hsize=6truein
\hoffset=.3truein
\vsize=8truein
\voffset=.3truein
\parskip=\medskipamount
\def\\{\cr}
\twelvepoint		
\doublespace		
\overfullrule=0pt	

\def\title			
  {\null\vskip 3pt plus 0.2fill
   \beginlinemode \doublespace \raggedcenter \bf}

\def\author			
  {\vskip 3pt plus 0.2fill \beginlinemode
   \singlespace \raggedcenter\sc}

\def\affil			
  {\vskip 3pt plus 0.1fill \beginlinemode
   \oneandahalfspace \raggedcenter \sl}

\def\abstract			
  {\vskip 3pt plus 0.3fill \beginparmode
   \oneandahalfspace ABSTRACT: }

\def\endpage{\vfill\eject}

\def\endtitlepage		
  {\endpage			
   \body}

\def\body			
  {\beginparmode}		

\def\head#1{			
  \goodbreak\vskip 0.5truein	
  {\immediate\write16{#1}
   \raggedcenter \uppercase{#1}\par}
   \nobreak\vskip 0.25truein\nobreak}

\def\references			
  {\head{References}		
   \beginparmode
   \frenchspacing \parindent=0pt \leftskip=1truecm
   \parskip=8pt plus 3pt \everypar{\hangindent=\parindent}}

\gdef\refis#1{\item{#1.\ }}			

\gdef\journal#1, #2, #3, 1#4#5#6{		
    {\sl #1~}{\bf #2}, #3 (1#4#5#6)}		

\def\pr{\journal Phys. Rev., }

\def\endreferences{\body}

\def\figurecaptions		
  {\endpage
   \beginparmode
   \head{Figure Captions}
}

\def\endpaper			
  {\endmode\vfill\supereject}
\def\endit
  {\endpaper\end}

\def\tag#1$${\eqno(#1)$$}

\def\align#1$${\eqalign{#1}$$}

\def\aligntag#1$${\gdef\tag##1\\{&(##1)\cr}\eqalignno{#1\\}$$
  \gdef\tag##1$${\eqno(##1)$$}}

\def\endaligntag{}

\def\overset #1\to#2{{\mathop{#2}\limits^{#1}}}
\def\underset#1\to#2{{\let\next=#1\mathpalette\undersetpalette#2}}
\def\undersetpalette#1#2{\vtop{\baselineskip0pt
\ialign{$\mathsurround=0pt #1\hfil##\hfil$\crcr#2\crcr\next\crcr}}}


\catcode`@=11
\newcount\tagnumber\tagnumber=0

\immediate\newwrite\eqnfile
\newif\if@qnfile\@qnfilefalse
\def\write@qn#1{}
\def\writenew@qn#1{}
\def\w@rnwrite#1{\write@qn{#1}\message{#1}}
\def\@rrwrite#1{\write@qn{#1}\errmessage{#1}}

\def\taghead#1{\gdef\t@ghead{#1}\global\tagnumber=0}
\def\t@ghead{}

\expandafter\def\csname @qnnum-3\endcsname
  {{\t@ghead\advance\tagnumber by -3\relax\number\tagnumber}}
\expandafter\def\csname @qnnum-2\endcsname
  {{\t@ghead\advance\tagnumber by -2\relax\number\tagnumber}}
\expandafter\def\csname @qnnum-1\endcsname
  {{\t@ghead\advance\tagnumber by -1\relax\number\tagnumber}}
\expandafter\def\csname @qnnum0\endcsname
  {\t@ghead\number\tagnumber}
\expandafter\def\csname @qnnum+1\endcsname
  {{\t@ghead\advance\tagnumber by 1\relax\number\tagnumber}}
\expandafter\def\csname @qnnum+2\endcsname
  {{\t@ghead\advance\tagnumber by 2\relax\number\tagnumber}}
\expandafter\def\csname @qnnum+3\endcsname
  {{\t@ghead\advance\tagnumber by 3\relax\number\tagnumber}}

\def\equationfile{%
  \@qnfiletrue\immediate\openout\eqnfile=\jobname.eqn%
  \def\write@qn##1{\if@qnfile\immediate\write\eqnfile{##1}\fi}
  \def\writenew@qn##1{\if@qnfile\immediate\write\eqnfile
    {\noexpand\tag{##1} = (\t@ghead\number\tagnumber)}\fi}
}

\def\callall#1{\xdef#1##1{#1{\noexpand\call{##1}}}}
\def\call#1{\each@rg\callr@nge{#1}}

\def\each@rg#1#2{{\let\thecsname=#1\expandafter\first@rg#2,\end,}}
\def\first@rg#1,{\thecsname{#1}\apply@rg}
\def\apply@rg#1,{\ifx\end#1\let\next=\relax%
\else,\thecsname{#1}\let\next=\apply@rg\fi\next}

\def\callr@nge#1{\calldor@nge#1-\end-}
\def\callr@ngeat#1\end-{#1}
\def\calldor@nge#1-#2-{\ifx\end#2\@qneatspace#1 %
  \else\calll@@p{#1}{#2}\callr@ngeat\fi}
\def\calll@@p#1#2{\ifnum#1>#2{\@rrwrite{Equation range #1-#2\space is bad.}
\errhelp{If you call a series of equations by the notation M-N, then M and
N must be integers, and N must be greater than or equal to M.}}\else%
 {\count0=#1\count1=#2\advance\count1
by1\relax\expandafter\@qncall\the\count0,%
  \loop\advance\count0 by1\relax%
    \ifnum\count0<\count1,\expandafter\@qncall\the\count0,%
  \repeat}\fi}

\def\@qneatspace#1#2 {\@qncall#1#2,}
\def\@qncall#1,{\ifunc@lled{#1}{\def\next{#1}\ifx\next\empty\else
  \w@rnwrite{Equation number \noexpand\(>>#1<<) has not been defined yet.}
  >>#1<<\fi}\else\csname @qnnum#1\endcsname\fi}

\let\eqnono=\eqno
\def\eqno(#1){\tag#1}
\def\tag#1$${\eqnono(\displayt@g#1 )$$}

\def\aligntag#1\endaligntag
  $${\gdef\tag##1\\{&(##1 )\cr}\eqalignno{#1\\}$$
  \gdef\tag##1$${\eqnono(\displayt@g##1 )$$}}

\def\eqalignno#1{\displ@y \tabskip\centering
  \halign to\displaywidth{\hfil$\displaystyle{##}$\tabskip\z@skip
    &$\displaystyle{{}##}$\hfil\tabskip\centering
    &\llap{$\displayt@gpar##$}\tabskip\z@skip\crcr
    #1\crcr}}

\def\displayt@gpar(#1){(\displayt@g#1 )}

\def\displayt@g#1 {\rm\ifunc@lled{#1}\global\advance\tagnumber by1
        {\def\next{#1}\ifx\next\empty\else\expandafter
        \xdef\csname @qnnum#1\endcsname{\t@ghead\number\tagnumber}\fi}%
  \writenew@qn{#1}\t@ghead\number\tagnumber\else
        {\edef\next{\t@ghead\number\tagnumber}%
        \expandafter\ifx\csname @qnnum#1\endcsname\next\else
        \w@rnwrite{Equation \noexpand\tag{#1} is a duplicate number.}\fi}%
  \csname @qnnum#1\endcsname\fi}

\def\ifunc@lled#1{\expandafter\ifx\csname @qnnum#1\endcsname\relax}

\let\@qnend=\end\gdef\end{\if@qnfile
\immediate\write16{Equation numbers written on []\jobname.EQN.}\fi\@qnend}

\catcode`@=12

\catcode`@=11
\newcount\r@fcount \r@fcount=0
\newcount\r@fcurr
\immediate\newwrite\reffile
\newif\ifr@ffile\r@ffilefalse
\def\w@rnwrite#1{\ifr@ffile\immediate\write\reffile{#1}\fi\message{#1}}

\def\writer@f#1>>{}
\def\referencefile{
  \r@ffiletrue\immediate\openout\reffile=\jobname.ref%
  \def\writer@f##1>>{\ifr@ffile\immediate\write\reffile%
    {\noexpand\refis{##1} = \csname r@fnum##1\endcsname = %
     \expandafter\expandafter\expandafter\strip@t\expandafter%
     \meaning\csname r@ftext\csname r@fnum##1\endcsname\endcsname}\fi}%
  \def\strip@t##1>>{}}

\def\citeall#1{\xdef#1##1{#1{\noexpand\cite{##1}}}}
\def\cite#1{\each@rg\citer@nge{#1}}	

\def\each@rg#1#2{{\let\thecsname=#1\expandafter\first@rg#2,\end,}}
\def\first@rg#1,{\thecsname{#1}\apply@rg}	
\def\apply@rg#1,{\ifx\end#1\let\next=\relax
\else,\thecsname{#1}\let\next=\apply@rg\fi\next}
\def\citer@nge#1{\citedor@nge#1-\end-}	
\def\citer@ngeat#1\end-{#1}
\def\citedor@nge#1-#2-{\ifx\end#2\r@featspace#1 
  \else\citel@@p{#1}{#2}\citer@ngeat\fi}	
\def\citel@@p#1#2{\ifnum#1>#2{\errmessage{Reference range #1-#2\space is bad.}%
    \errhelp{If you cite a series of references by the notation M-N, then M and
    N must be integers, and N must be greater than or equal to M.}}\else%
 {\count0=#1\count1=#2\advance\count1
by1\relax\expandafter\r@fcite\the\count0,%
  \loop\advance\count0 by1\relax
    \ifnum\count0<\count1,\expandafter\r@fcite\the\count0,%
  \repeat}\fi}

\def\r@featspace#1#2 {\r@fcite#1#2,}	
\def\r@fcite#1,{\ifuncit@d{#1}
    \newr@f{#1}%
    \expandafter\gdef\csname r@ftext\number\r@fcount\endcsname%
                     {\message{Reference #1 to be supplied.}%
                      \writer@f#1>>#1 to be supplied.\par}%
 \fi%
 \csname r@fnum#1\endcsname}
\def\ifuncit@d#1{\expandafter\ifx\csname r@fnum#1\endcsname\relax}%
\def\newr@f#1{\global\advance\r@fcount by1%
    \expandafter\xdef\csname r@fnum#1\endcsname{\number\r@fcount}}

\let\r@fis=\refis			
\def\refis#1#2#3\par{\ifuncit@d{#1}
   \newr@f{#1}%
   \w@rnwrite{Reference #1=\number\r@fcount\space is not cited up to now.}\fi%
  \expandafter\gdef\csname r@ftext\csname r@fnum#1\endcsname\endcsname%
  {\writer@f#1>>#2#3\par}}

\def\ignoreuncited{
   \def\refis##1##2##3\par{\ifuncit@d{##1}%
     \else\expandafter\gdef\csname r@ftext\csname
r@fnum##1\endcsname\endcsname%
     {\writer@f##1>>##2##3\par}\fi}}

\def\r@ferr{\endreferences\errmessage{I was expecting to see
\noexpand\endreferences before now;  I have inserted it here.}}
\let\r@ferences=\references
\def\references{\r@ferences\def\endmode{\r@ferr\par\endgroup}}
\let\endr@ferences=\endreferences
\def\endreferences{\r@fcurr=0
  {\loop\ifnum\r@fcurr<\r@fcount
    \advance\r@fcurr by 1\relax\expandafter\r@fis\expandafter{\number\r@fcurr}%
    \csname r@ftext\number\r@fcurr\endcsname%
  \repeat}\gdef\r@ferr{}\endr@ferences}

\let\r@fend=\endpaper\gdef\endpaper{\ifr@ffile
\immediate\write16{Cross References written on []\jobname.REF.}\fi\r@fend}

\catcode`@=12

\def\(#1){(\call{#1})}
\def\ssc{\scriptscriptstyle}

\def\-#1{_{\ssc {#1} }}

\def\hat #1{\mathaccent94{#1}}


\title THERMAL EQUILIBRIUM FROM THE HU-PAZ-ZHANG MASTER EQUATION
\author J.R. Anglin\footnote{$^\dagger$}{anglin@hep.physics.mcgill.ca}
\affil Physics Department, McGill University
3600 University Street
Montr\'eal, Qu\'ebec CANADA H3A 2T8
\abstract The exact master equation for a  harmonic oscillator coupled to a
heat bath,  derived recently by Hu, Paz and Zhang, is simplified by taking the
weak-coupling, late-time limit.  The unique time-independent solution to this
simplified master equation is the canonical ensemble at the temperature of the
bath.  The frequency of the oscillator is effectively lowered by the
interaction with the bath.

\endtitlepage
\eject

\head{Introduction}

The evolution of the density matrix of a quantum harmonic oscillator linearly
coupled to a heat bath is a fundamental problem, but it has only recently been
solved exactly[\cite{HPZ}].
Physical quantum mechanical systems have often been idealized as isolated,
with the sole effect of the environment being the maintenance of a finite
temperature.  The non-trivial role of the environment in decohering the
excitations of a weakly-coupled system is currently still being explored; as a
complement to this research, this letter examines the ground state of a
harmonic oscillator weakly coupled to a heat bath.

If a system is in equilibrium with a thermal environment, to which it is weakly
coupled, it has long been assumed that the reduced density matrix for the
system
is given by the canonical ensemble at the environmental temperature.
Using the density matrix evolution equation (`master equation') derived in
Reference [\cite{HPZ}], this letter confirms this
assumption in the case of the harmonic oscillator.  The discussion assumes
familiarity with the results of Reference [\cite{HPZ}] (hereafter denoted HPZ),
which will not be derived here.

\head{The HPZ master equation at weak coupling and late time}

In HPZ, the following master equation is derived, for the
time evolution of the density matrix $\rho(x,x')$ of a simple harmonic
oscillator, with the position variable $x$ coupled linearly to a heat bath:
$$\eqalign{
i{\partial\over\partial t}\rho =& -{1\over2}
\Bigl[{\partial^2\over\partial x^2}-{\partial^2\over\partial x'^2}
                         - \Omega\-0^2(x^2-x'^2)\Bigr]\rho\cr
&+\,(x-x')\Bigl[A(t)({\partial\over\partial x}+{\partial\over\partial x'}) +
                                 B(t)(x+x')\Bigr]\rho\cr
&-i(x-x')\Bigl[C(t)({\partial\over\partial x}-{\partial\over\partial x'})
             + D(t)(x-x')\Bigr]\rho\;.}\eqno(tdme)
$$
Here $\hbar$ and the oscillator mass have been set equal
to 1.  The time-dependent real co-efficients defined in HPZ have been
renamed $A$, $B$, $C$, and $D$.  The terms in \(tdme) proportional to $A$ and
$D$ are
responsible for diffusive effects in the evolution of $\rho$.
$B$ could be considered a time-dependent addition to the
effective frequency of the oscillator, while $C$ has the effect of a
time-dependent dissipation constant.

Equation \(tdme) is derived using an initial quantum state that is a direct
product of the initial oscillator and (thermal) environment states; the authors
of HPZ suggest that some features of their master equation may be artifacts of
this actually rather implausible initial condition. In order to avoid the
spurious time dependences introduced by the artificial  assumption that the
oscillator and heat bath are uncorrelated at time $t=0$, the co-efficients will
be replaced by their asymptotic forms at late times: $A(t)\to A\-\infty\equiv
A(\infty)$, etc.  Setting the RHS of \(tdme) equal to zero, one obtains an
equation for the time-independent state into which the open harmonic oscillator
might be expected to settle down at late time.  Equivalently, this equation can
be considered to describe the ground state of a simple harmonic oscillator in
the presence of a thermal environment:
$$
\eqalign{0\;=\;&
-{1\over2}\Bigl[{\partial^2\over\partial x^2}-{\partial^2\over\partial x'^2}
           - \Omega\-0^2(x^2-x'^2)\Bigr]\rho\cr
&+(x-x')\Bigl[A\-\infty ({\partial\over\partial x}+{\partial\over\partial x'})
           + B\-\infty (x+x')\Bigr]\rho\cr
&-i(x-x')\Bigl[C\-\infty ({\partial\over\partial x}-{\partial\over\partial x'})
           + D\-\infty (x-x')\Bigr]\rho\;.}\eqno(time)
$$

The time-independent co-efficients have simple forms in the weak-coupling
limit, where only terms of up to second order in the bath-oscillator coupling
constant are retained[HPZ, eq. 2.46].  In the notation of the present letter,
$$\eqalign{
A\-\infty =& g^2{1\over\Omega\-0}\lim_{t\to\infty}\int_0^t\!ds\!\int_0^\Gamma\!
    	d\omega\,I(\omega)\coth{\beta\omega\over2}
    	\cos\omega s \sin\Omega\-0 s\cr
B\-\infty =&
-g^2\lim_{t\to\infty}\int_0^t\!ds\!\int_0^\Gamma\!d\omega\,I(\omega)
       	\sin\omega s  \cos\Omega\-0 s\cr
C\-\infty =& g^2{1\over\Omega\-0}\lim_{t\to\infty}\int_0^t\!ds\!
   	\int_0^\Gamma\!d\omega\,I(\omega)
      	\sin\omega s \sin\Omega\-0 s\cr
D\-\infty =& g^2\lim_{t\to\infty}\int_0^t\!ds\!\int_0^\Gamma\!d\omega\,
   	I(\omega)\coth{\beta\omega\over2}
        \cos\omega s \cos\Omega\-0 s\;.}\eqno(coef)
$$
$I(\omega)$ is the spectral density of the environmental heat bath, and $g$
is the bath-oscillator coupling constant.  It is assumed that
$I(\omega)$ is cut off at some high frequency $\Gamma$ and vanishes
at least linearly at zero frequency, so
that, as long as the limit
$t\to\infty$ is taken last, the order of the $s$ and $\omega$ integrals is
arbitrary.  The temperature of the bath is $kT=\beta^{-1}$, where $k$ is the
Boltzman constant.

In solving \(time) to first order in $g^2$, the co-efficient
$B\-\infty$ will be split into two terms:
$$
B\-\infty \equiv \Omega\-0\tanh{\beta\Omega\-0\over2} A\-\infty +
\Omega\-0\delta\Omega\;.\eqno(bsplit)
$$
The first term is chosen to provide a cancellation shown below, while the
second term is a renormalizing correction to the effective frequency of the
oscillator, proportional to $g^2$.  ($\delta\Omega$ will be determined in the
next section of this letter.)  Define the renormalized frequency to be
$\Omega\equiv \Omega\-0 + \delta\Omega$.

A standard representation of the delta function allows one to write
$$\eqalign{
C\-\infty &= {g^2\pi \over2\Omega\-0}\,I(\Omega\-0)
\simeq {g^2\pi \over2\Omega}\,I(\Omega)\cr
D\-\infty &= {g^2\pi \over2}\coth{\beta\Omega\-0\over2}\,I(\Omega\-0)
\simeq {g^2\pi \over2}\coth{\beta\Omega\over2}\,I(\Omega)\;.}\eqno(CD)
$$
One can therefore re-write \(time) in the final form
$$
\eqalign{0\;=\;&
-{1\over2}\Bigl[{\partial^2\over\partial x^2}-{\partial^2\over\partial x'^2}
           - \Omega^2(x^2-x'^2)\Bigr]\rho\cr
&+(x-x')A\-\infty\Bigl[ ({\partial\over\partial x}+{\partial\over\partial x'})
           +\Omega\tanh{\beta\Omega\over2}(x+x')\Bigr]\rho\cr
&-i(x-x'){g^2\pi \over2\Omega}I(\Omega)
           \Bigl[({\partial\over\partial x}-{\partial\over\partial x'})
           + \Omega\coth{\beta\Omega\over2} (x-x')\Bigr]\rho\;,}\eqno(me)
$$
keeping only terms of up to first order in $g^2$.  This is the weak-coupling,
infinite-time limit of the exact time-independent master equation of HPZ.

Define the thermal density matrix
$$
\rho\-\beta(x,x') = (1-e^{-\beta\Omega})\sum_{n=0}^\infty
e^{-n\beta\Omega}\,\psi\-n(x)\psi\-n(x')\;,\eqno(can)
$$
where $\psi\-n(x)$ is the wave function for the $n$th excited state of the
harmonic oscillator with frequency $\Omega$.  It will now be verified
that $\rho = \rho\-\beta$ is a solution to \(me).  The verification is
straightforward, and proceeds line by line.

The first line of \(me) is simply $[\hat H_x -\hat H_{x'}]\rho\-\beta$,
for $\hat H$ the
harmonic oscillator Hamiltonian in the position representation, and so
equals zero.

The second line is proportional to
$$\eqalign{
&\sum_{n=0}^\infty e^{-n\beta\Omega}\Bigl[\Omega^{1/2}\sinh{\beta\Omega\over2}
(x + x') + \Omega^{-1/2}\cosh{\beta\Omega\over2} ({\partial\over\partial x} +
{\partial\over\partial x'})
\Bigr]\psi\-n(x)\psi\-n(x')\cr
=&{1\over\sqrt2}\sum_{n=0}^\infty e^{-n\beta\Omega}\Bigl[e^{\beta\Omega\over2}
(\hat a + \hat a') - e^{-{\beta\Omega\over2}}(\hat a^\dagger
   + \hat a'^\dagger)\Bigr]\psi\-n(x)\psi\-n(x')\cr
=&{e^{\beta\Omega\over2}\over\sqrt2}\sum_{n=0}^\infty e^{-n\beta\Omega}
\Bigl[\sqrt{n}\,\psi\-{n-1}(x)\psi\-n(x')
  -e^{-\beta\Omega}\sqrt{n+1}\,\psi\-{n+1}(x)\psi\-n(x')
   + [x\leftrightarrow x']\Bigr]\cr
=&{e^{-{\beta\Omega\over2}}\over\sqrt2}
\sum_{n=0}^\infty e^{-n\beta\Omega} \sqrt{n+1}\Bigl[
\psi\-n(x)\psi\-{n+1}(x') - \psi\-{n+1}(x)\psi\-n(x') + [x\leftrightarrow x']
\Bigr]\cr
=&\ 0\;\;.}\eqno(l2van)
$$
Here $\hat a = {1\over\sqrt{2\Omega}}(\Omega x + {\partial\over\partial x})$,
$\hat{a}^\dagger =
{1\over\sqrt{2\Omega}}(\Omega x - {\partial\over\partial x})$, the primes
imply that $x$ is replaced by $x'$, and $\psi\-n$ satisfies
$$\eqalign{
\hat a \psi\-n &= \sqrt{n}\,\psi\-{n-1}\cr
\hat a^\dagger \psi\-n &=\sqrt{n+1}\,\psi\-{n+1}\cr
\psi\-{-1}&\equiv 0\;.}
$$

Similarly, the third line of \(me) is proportional to
$$\eqalign{
 &\sum_{n=0}^\infty e^{-n\beta\Omega}\Bigl[\Omega^{1/2}\cosh{\beta\Omega\over2}
        (x-x') +\Omega^{-1/2}\sinh{\beta\Omega\over2}
	({\partial\over\partial x}-{\partial\over\partial x'})\Bigr]
	\psi\-n(x)\psi\-n(x')\cr
=&{1\over\sqrt2}\sum_{n=0}^\infty e^{-n\beta\Omega}\Bigl[e^{\beta\Omega\over2}
	(\hat a - \hat a') + e^{-{\beta\Omega\over2}}(\hat{a}^\dagger
	-\hat{a}'^\dagger)\Bigr]\psi\-n(x)\psi\-n(x')\cr
=&{e^{\beta\Omega\over2}\over\sqrt2}\sum_{n=0}^\infty e^{-n\beta\Omega}
	\Bigl[\sqrt{n}\,\psi\-{n-1}(x)\psi\-n(x') +
	e^{-\beta\Omega}\sqrt{n+1}\,\psi\-{n+1}(x)\psi\-n(x')
        - [x\leftrightarrow x']\Bigr]\cr
=&{e^{-{\beta\Omega\over2}}\over\sqrt2}\sum_{n=0}^\infty e^{-n\beta\Omega}
	\sqrt{n+1}\Bigl[\psi\-n(x)\psi\-{n+1}(x') +
	\psi\-{n+1}(x)\psi\-n(x') - [x\leftrightarrow x']\Bigr]\cr
=&\ 0 \; \; . } \eqno(l3van)
$$

The canonical ensemble at
temperature $kT=\beta^{-1}$ is therefore a solution to the late-time, weak
coupling,
time-independent master equation.  This equation is a hyperbolic
partial differential equation in the two independent variables $x$ and $x'$.
Normalizability of $\rho$ requires that $\rho$ and its derivatives decay to
zero
for large $|x|$ or $|x'|$.  Normalizability therefore imposes Cauchy initial,
final, and boundary conditions
on the master equation, which in general over-determine the
solution[\cite{mf1}].  The
solution which has been found is therefore unique.  (The terms
``initial'' and ``final'' are used by analogy, since either $x$ or $x'$
may be considered to play the
role of the timelike variable usually involved in hyperbolic equations.)

\head{Determination of $\delta\Omega$}

{}From equations \(coef) and \(bsplit) one can determine the frequency
correction
$\delta\Omega$.  Comparing the definitions of $C\-\infty$ and $D\-\infty$, one
sees that
$$
\delta\Omega = -{g^2\over\Omega}\lim_{t\to\infty}\int_0^\Gamma\!d\omega\,
    I(\omega)\coth{\beta\omega\over2}\, R(\omega,\Omega,t)\;,\eqno(om1)
$$
where terms of order $g^4$ are neglected, and $R$ is defined by
$$
R(\omega,\Omega,t) \equiv
\tanh{\beta\omega\over2}\int_0^t\!ds\,\sin\omega s\cos\Omega s
+ \tanh{\beta\Omega\over2}\int_0^t\!ds\,\cos\omega s \sin\Omega s\;.\eqno(R)
$$
$R$ can be evaluated by expanding the hyperbolic tangents in Taylor series, and
noting that, for integer $n>0$,
$$\eqalign{R\-n(\omega,\Omega,t)&\equiv
\omega^{2n+1}\int_0^t\!ds\,\sin\omega s\cos\Omega s +
\Omega^{2n+1}\int_0^t\!ds\,\cos\omega s\sin\Omega s\cr
&={1\over\omega^2-\Omega^2}\Bigl[(\omega^{2(n+1)}-\Omega^{2(n+1)})(1-
\cos\omega t\cos\Omega )\cr
&\qquad\qquad\qquad\qquad
 - \omega\Omega(\omega^{2n}-\Omega^{2n})\sin\omega t\sin\Omega
t\Bigr]\cr
&= {\omega^{2(n+1)}-\Omega^{2(n+1)}\over\omega^2-\Omega^2} - \sum_{m=0}^n
\omega^{2m}\Omega^{2(n-m)}\cos\omega t\cos\Omega t\cr
&\qquad\qquad\qquad\qquad -
\omega\Omega\sum_{m=0}^{n-1}\omega^{2m}\Omega^{2(n-1-m)}\sin\omega t\sin\Omega
t\;.}\eqno(deldel)
$$
One then has
$$
\delta\Omega = -{g^2\over\Omega}\lim_{t\to\infty}\sum_{n=0}^\infty
A\-n\Bigl({\beta\over2}\Bigr)^{2n+1}
\int_0^\Gamma\!d\omega\,I(\omega)\coth{\beta\omega\over2}\,
R\-n(\omega,\Omega,t)\;,
\eqno(Rn)
$$
where $A\-n$ are the co-efficients in the Taylor series for the hyperbolic
tangent.

The oscillating functions of $t$ in the last two lines of \(deldel) may in fact
be ignored in the limit $t\to\infty$: since $I(\omega)$ vanishes (at least)
linearly at $\omega = 0$, $I(\omega)\coth{\beta\omega\over2}$ may be expanded
in a series of non-negative powers of $\omega$, and explicit calculation will
show that
$$
\lim_{t\to\infty}\int_0^\Gamma\!dx\, x^m e^{itx}=0\;\eqno(vanvan)
$$
for $m\geq 0$.

Ignoring the $t$-dependent part of $R\-n(\omega,\Omega,t)$, one therefore has,
to leading order in $g^2$,
$$
\delta\Omega = -{g^2\over\Omega}
\int_0^\Gamma\!d\omega\,
I(\omega)\coth{\beta\omega\over2}\,\left[{ {\omega \tanh{\beta\omega\over2}
-\Omega\tanh{\beta\Omega\over2}}\over{\omega^2-\Omega^2}
}\right]\;.\eqno(deltao)
$$
In the high temperature limit, this becomes
$$
\delta\Omega\left.\right\vert_{\beta\to zero}
=-{g^2\over\Omega}\int_0^\Gamma\!d\omega\,{I(\omega)\over\omega}\;,
\eqno(highT)
$$
while in the low temperature limit it approaches
$$
\delta\Omega\left.\right\vert_{\beta\to\infty}
=-{g^2\over\Omega}\int_0^\Gamma\!d\omega\,
{I(\omega)\over\omega+\Omega}\;.
\eqno(lowT)
$$

\head{Conclusion}

In recent years it has been argued that open quantum systems are generically
decohered by the environment to which they are coupled, in such a way that the
state of the system is rapidly driven towards a mixture of eigenstates of the
interaction Hamiltonian[\cite{HPZ, zurek, calleg, unzur}].  This
phenomenon is considered to occur on a short time scale; open quantum systems
in
equilibrium after a long period of time have received comparatively little
attention.  It is nevertheless of fundamental importance to
confirm that the equilibrium state of a system whose position variable is
weakly coupled to a  heat bath is indeed the canonical ensemble, and not a
mixture of position eigenstates.  This is true at all temperatures, and is
independent of the spectral density of the heat bath, to leading order in the
coupling $g^2$.

At higher order in $g^2$, this result probably no longer holds.  Heuristic
arguments justifying the canonical ensemble typically assume weak coupling
between the system and its environment.  Presumably the equilibrium solution
for $\rho$ depends at higher orders in $g^2$ on the specific form of the
spectral density $I(\omega)$.

The infinite-time limit of the master equation derived using uncorrelated
initial states at $t=0$ may be conjectured to be equivalent to the master
equation one would obtain at finite times using more realistic initial
conditions.  The
canonical ensemble is an explicit example of a state which does not (at leading
order in the coupling constant) suffer wave-function collapse onto the pointer
basis determined by the coupling to the environment.  This supports the
suggestions by several previous researchers that the role of the initial
conditions in decoherence and environmentally-induced superselection needs
further investigation.

\head{Acknowledgements}

The author gratefully acknowledges valuable conversations with R.C. Myers.
This research was supported by NSERC of Canada, and by the Fonds pour la
Formation de Chercheurs et l'Aide \`a la Recherche du Qu\'ebec.

\references
\singlespace

\refis{HPZ} B.L.~Hu, Juan~Pablo~Paz, and Yuhong~Zhang, \pr D45, 2843, 1992.

\refis{zurek} W.H.~Zurek, \pr D26, 1862, 1982.

\refis{unzur} W.G.~Unruh and W.H.~Zurek, \pr D40, 1071, 1989.

\refis{calleg} A.O.~Caldeira and A.J.~Leggett, \pr A31, 1059, 1985.

\refis{mf1} P.M. Morse and H. Feshbach, {\it Mathematical Methods of
Theoretical
Physics} (McGraw-Hill; New York, 1953), pp. 683ff.

\endreferences

\endit